\documentclass[preprint,reprint,amsmath,showpacs,amssymb,superscriptaddress]{revtex4-1}

\usepackage{graphicx}
\usepackage{dcolumn}
\usepackage{bm}
\usepackage{color}
\begin{document}
\title{Scaling of the Kondo zero bias peak in a hole quantum dot at finite temperatures}

\author{O. Klochan}
\email{klochan@phys.unsw.edu.au}
\author{A.P. Micolich}
\author{A.R. Hamilton}
\affiliation{School of Physics, University of New South Wales,
Sydney NSW 2052, Australia.}

\author{D. Reuter}
\author{A. D.  Wieck}
\affiliation{Angewandte Festk\"{o}rperphysik, Ruhr-Universit\"{a}t Bochum, D-44780 Bochum,
Germany}

\author{F. Reininghaus}
\author{M. Pletyukhov}
\author{H. Schoeller}
\affiliation{Institut f\"{u}r Theorie der Statistischen Physik and JARA - Fundamentals of
Future Information Technology, RWTH Aachen, 52056 Aachen, Germany }

\date{\today}

\begin{abstract}

We have measured the zero bias peak in differential conductance in a hole quantum dot. We have scaled the experimental data with applied bias and compared to real time renormalization group calculations of the differential conductance as a function of source-drain bias in the limit of zero temperature and at finite temperatures. The experimental data show  deviations from the $T=0$ calculations at low bias, but are in very good agreement with the finite $T$ calculations. The Kondo temperature $T_K$ extracted from the data using $T=0$ calculations, and from the peak width at $\frac{2}{3}$ maximum, is significantly higher than that obtained from finite $T$ calculations.
\end{abstract}

\maketitle

The Kondo effect arises due to interaction of a single localized electron spin with a sea of delocalized electron spins~\cite{Kondo64}, and it was first  observed through a nonmonotonic temperature dependence of the resistivity of a metal doped with magnetic impurities~\cite{Haas33}. The strength of this interaction is characterized by a single parameter called the Kondo temperature $T_K$. In metals the Kondo effect is due to the contribution of many magnetic impurities, so it is interesting to ask: What happens in the case of a single impurity? This question was  answered experimentally in low temperature STM experiments \cite{MadhavanScience98} where individual Co atoms were placed at Au surface and were probed with an STM tip. Alternatively, a quantum dot can be used to study  the Kondo effect~\cite{GoldhaberNat98, CronenwettScience98}. Here, a single unpaired electron spin trapped in the dot interacts with spins of electrons in the source and drain resulting in  a many-body state which enhances conductance through the dot at low temperature. The major advantage of a quantum dot is the ability to tune the Kondo temperature over a wide range by changing a bias on one of the gate electrodes forming the dot~\cite{KouwenhovenPhysWorld}.

The usual method for measuring $T_K$ in a quantum dot is to study the temperature dependence of the conductance and compare to theory using an empirical expression with $T_K$ as a fitting parameter~\cite{GoldhaberPRL98}. Recently a much simpler technique has become available based on analysing the conductance as a function of source-drain bias instead of temperature. The Kondo enhanced conductance is suppressed by a source drain bias $V_{SD}$ across the dot, producing a characteristic peak in the differential conductance $G^{\prime}(V_{SD})$ -  the zero bias peak (ZBP). Recently~\cite{PletyukhovPRL12}, a two-loop real time renormalization group theory  (RTRG) has been applied to  a Kondo-$\frac{1}{2}$ model and provided numerical results for the bias dependence of differential conductance in the limit of zero temperature $T = 0$. These calculations have been used in recent experiments on InAs nanowires~\cite{KretininPRB12} to extract values of $T_K$ and compare to the values obtained from the  linear conductance measured as a function of temperature.  Here we measure the zero bias peak in differential conductance in a spin-$\frac{3}{2}$ hole quantum dot. We scale the zero bias peak  as a function of $V_{SD}$ and compare to  RTRG calculations in the limit of $T=0$ and at finite $T$. We observe  deviations between our experimental data and the $T=0$ calculations~\cite{PletyukhovPRL12} at low bias. We show that these deviations arise from the  finite measurement temperature and can be taken into account by scaling the experimental data to newly available finite $T$ calculations~\cite{Reininghaus12}. These results suggest that the  Kondo effect in a spin-$\frac{3}{2}$ GaAs hole quantum dot can be accurately described by  a Kondo-$\frac{1}{2}$ model.

 The system studied here is a hole quantum dot defined in the two-dimensional (2D) hole system at a GaAs-AlGaAs heterojunction. It is technically difficult to fabricate small hole quantum dots to access the single hole regime due to the large hole effective mass ($m_h = 0.2-0.5\times m_0$) and poor electrical stability of hole nanodevices~\cite{EnsslinNP06}. Instead, we form a very small  quantum dot in a quantum wire near pinch-off due to the roughness of the wet etching used to define the wire (see e.g.~Ref.~\cite{SfigakisPRL08}). The wire is etched $20$~nm deep through degenerately doped cap layer used to induce the carriers electrostatically, which results in rather stable devices~\cite{KlochanAPL06}. Details of the fabrication can be found in Ref.~\cite{KlochanPRL11}. All measurements  were taken in a dilution refrigerator at a top gate voltage $V_{TG}=-0.67$~V corresponding to a two dimensional density $p=2\times10^{11}$~cm$^{-2}$  and mobility $\mu=450,000$~cm$^2$/Vs.

It is not obvious that Kondo physics will be the same for holes as for spin-$\frac{1}{2}$ electrons, since holes in GaAs have a $p$-type wave function and non zero orbital momentum $l$, giving a total angular momentum $j = l+s = \frac{3}{2}$. Holes have much stronger spin-orbit coupling, and very different spin properties, to electrons: In a two dimensional hole system the fourfold degeneracy of the hole bands is split into two doubly degenerate bands with projections of $j =\pm \frac{3}{2}$  (heavy holes) and $j =\pm \frac{1}{2}$ (light holes). We work in the low carrier density regime where only the heavy hole band is occupied. The holes are then confined to lower dimensions using surface gates, which give a weaker confinement potential than the self-consistent triangular well at the heterointerface. This suggests that the nature of the hole states remains predominantly heavy hole like. This is consistent with the anisotropy of  the splitting of the one-dimensional subbands with in-plane field~\cite{ChenNJP10} being the same as the anisotropic response of the ZBP~\cite{KlochanPRL11}. Therefore it is possible a Kondo-$\frac{1}{2}$ model can be applied, since the Kondo effect only requires a doubly degenerate level to facilitate transport, independent of the exact nature of the spin or pseudospin.

Figure~1(a) shows  the differential conductance $G^{\prime}$ as a function of applied source drain bias~\cite{bias}  $V_{SD}$ measured for a series of side-gate voltages $V_{SG}$ corresponding to $10^{-3} < G^{\prime}(V_{SD}=0) < 0.4 \times 2e^{2}/h$. A clear  zero bias peak centered at $V_{SD}=0$~V is present in all traces. It is important to note that the zero bias peak reported here is distinct from the zero bias anomaly observed in quantum wires. In quantum wires the zero bias anomaly shows a linear splitting in magnetic field that is exponentially dependent on $V_{SG}$~\cite{CronenwettPRL02,SarkozyPRB09}. The zero bias anomaly in electron quantum dots, and in the hole dot studied here, also splits with an applied magnetic field, but the splitting is independent of $V_{SG}$~\cite{GoldhaberNat98,CronenwettScience98,KlochanPRL11}

\begin{figure}
\includegraphics[width = 8.5cm]{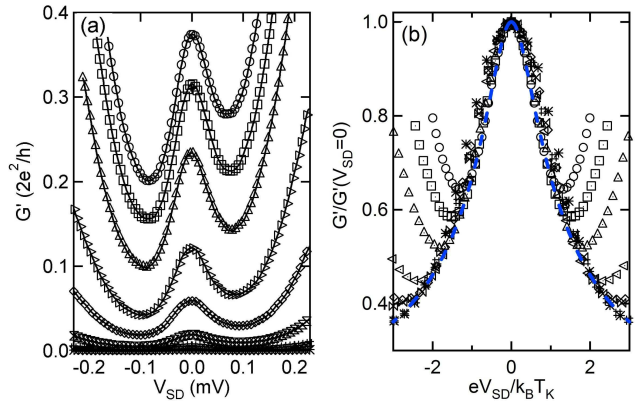}
\caption{\label{fig1} (Color online) (a) Differential conductance $G^{\prime}$ below $0.4\times2e^2/h$ as a function of $V_{SD}$ showing a pronounced ZBP for all traces. Top trace is measured at $V_{SG}=1.08$~V,  bottom trace at $V_{SG}=1.115$~V and the step size between  adjacent traces is $5$~mV. (b) Experimental data (symbols) from Fig.~1a symmetrized and scaled  onto the universal trace for $T=0$ (dashed blue line). For clarity only every 10$^{th}$ experimental data point is shown. }
\end{figure}

Observation of the signatures of the Kondo effect in magnetic field~\cite{KlochanPRL11} indicates that the dot is small enough to see  Kondo physics, and that we sit in a single Kondo valley where the dot occupancy is odd~\cite{odd}. The zero bias peak in Fig.~1(a) is superimposed on a rising background, which has been attributed to charge fluctuations  in the mixed-valence regime in quantum dots \cite{GoldhaberPRL98,KretininPRB11}, or enhanced tunneling through a single barrier~\cite{MartinMorenoJPCM92}. The conductance traces are asymmetric with respect to $V_{SD}$ due to self-gating~\cite{MartinMorenoJPCM92}, which  we eliminate in our analysis by symmetrizing the experimental data with respect to $V_{SD} = 0$~V. We can estimate the Kondo temperature from the width of the zero bias peak. The $T=0$ theory~\cite{PletyukhovPRL12, KretininPRB12} gives the full width at $\frac{2}{3}$ maximum as FW2/3M$=2T_K$. As we go down from the top trace to the bottom in Fig.~1(a), the conductance drops by almost three orders of magnitude whereas the FW2/3M does not change significantly. Because  $T_K$ depends exponentially on the coupling between the dot and the leads  it should change drastically \cite{GoldhaberPRL98}. This implies that the dot is coupled asymmetrically to the leads with the more opaque barrier controlling  the overall conductance while the more transparent barrier defines the Kondo temperature~\cite{PustilnikJPC04}. When the conductance exceeds $e^2/h$ one of the barriers becomes completely transparent and we measure transport through  a single barrier only, therefore to probe quantum dot physics we need to stay in the low conductance regime below $e^2/h$. Thus the Kondo enhancement in these measurements cannot reach the unitary conductance limit $G_0 = 2e^2/h$.


 We  scale our data  to numerically calculated $G^{\prime}(V_{SD})$  in the limit of $T=0$ using  $T_K$ as a single fitting parameter~\cite{PletyukhovPRL12}, following a similar procedure to Ref.~\cite{KretininPRB12}.  To account for the asymmetric barriers we first normalize each ZBP trace by its zero bias value $G^{\prime}(V_{SD}=0)$  to obtain $G^{\prime}/G^{\prime}(V_{SD}=0)$ (even when the system is asymmetric this procedure is still valid as shown previously~\cite{KretininPRB12, PletyukhovPRL12}, since the fit is primarily driven by the  steepness of the conductance drop with applied source drain bias). Second, we plot both the experimental and the calculated $G^{\prime}(V_{SD})/G^{\prime}(V_{SD}=0)$ as a function of scaled bias $eV_{SD}/k_BT_K$ on a semi-log scale. Third, we shift each experimental trace along the $eV_{SD}/k_BT_K$-axis (adjust $T_K$) until it overlaps with the theoretical curve (because the ZBP is superimposed onto the rising background, the experimental data should always be above the calculations, particularly at large biases). The value of the shift required to make  the experimental and calculated traces overlap then gives $T_K$.

\begin{figure}
\includegraphics[width = 8.5cm]{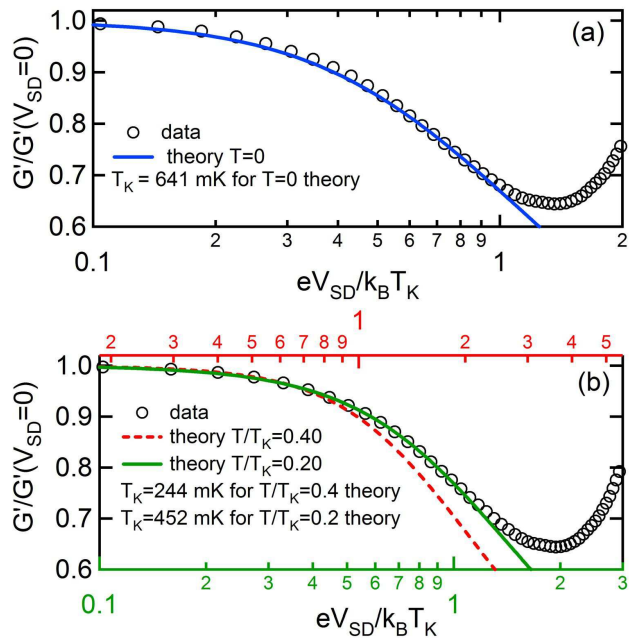}
\caption{\label{fig2} (Color online) (a) An experimental trace at $V_{SG}=1.08$~V (symbols) compared with the calculated traces for  $T/T_K=0$ (solid blue line). (b) The same experimental data compared with finite temperature calculations for $T/T_K=0.20$ (green solid line, bottom x-axis) and $T/T_K=0.4$ (red dashed line, top x-axis).  For clarity only every 5$^{th}$ experimental data point is shown.}
\end{figure}

The results of scaling are presented in Fig.~1(b). Note that although the scaling is performed on a semi-log scale, the scaled data is shown on a linear scale to highlight the ZBP. At first sight the scaling of the experimental data (black symbols) to the calculated trace for $T=0$  (blue dashed line) looks very good, apart from the upturn due to the rising  background in the experimental data at high bias ($eV_{SD}/k_BT_K>1$).  However, closer inspection of Fig.~1(b) reveals that the experimental data are consistently above the $T=0$ calculations in the low bias window $|eV_{SD}/k_BT_K|<1$, where the agreement between theory and experiment should be best. An example of this is shown in Fig.~2(a), where a single experimental trace measured for  $V_{SG}=1.08$~V (the top trace in Fig.~1(a)) is fitted to the $T=0$ theory using $T_K$ as the only adjustable parameter.  We can see that the experimental data lie above the theory for most of the low bias region. The fact that our data in Fig.~2(a) deviate from the calculated trace for $T=0$  suggests that the measurements are not in the low $T/T_K$ limit. Previous experiments~\cite{KretininPRB11} which showed good agreement between the $T=0$ theory and experiment were performed in a system with $T/T_K < 0.03$. In contrast, we performed experiments on a system with smaller $T_K$  and higher  measurement temperature so that we can reach higher $T/T_K$.

\begin{figure}
\includegraphics[width = 8.5cm]{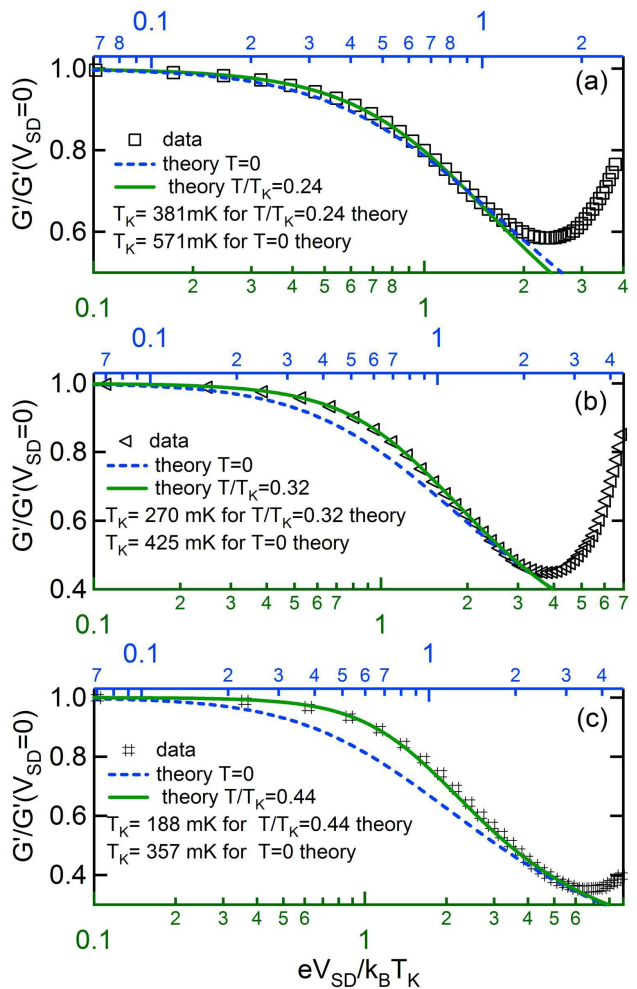}
\caption{\label{fig3} (Color online)  Experimental conductance data (symbols) compared with theoretical fits for $T/T_K=0$ (blue dashed line, top x-axis) and  $T/T_K >0$ (green solid line, bottom x-axis) in the low bias regime.  (a) $V_{SG}= 1.085$~V , (b) $V_{SG}= 1.095$~V and (c) $V_{SG}= 1.11$~V.  For clarity only every 5$^{th}$ experimental  data point is shown. }
\end{figure}

The RTRG calculations of the differential conductance $G^{\prime}(eV_{SD}/k_BT_K,T/T_K)$ in Ref.~\cite{PletyukhovPRL12} only consider the limiting cases where either the bias voltage $V_{SD}$ or the temperature $T$ are zero. Very recently~\cite{Reininghaus12} it has become possible to extend these  calculations so that $G^{\prime}(eV_{SD}/k_BT_K,T/T_K)$ can be computed for all $V_{SD}$ and $T$.  We have used 21 numerically calculated traces of $G^{\prime}(eV_{SD}/k_BT_K,T/T_K)$ in the range $0<T/T_K<0.8$ in steps of $0.04$ to find the best fit to each of our experimental traces. To demonstrate the improved agreement between the finite $T$ theory and the experimental data we show different calculated traces fitted to a single experimental trace in Fig.~2(b). We recall that the calculations for $T/T_K=0$ in Fig.~2(a) lie below the experimental data, because the ratio $T/T_K=0$ is too small to produce a good fit -- on the semi-log axes of Fig~2(a) changing $T_K$ only shifts the experimental data sideways, without altering the slope of the curve. In Fig.~2(b) the same experimental data are plotted against both the top (red) the bottom (green) x-axis, along with the optimal theory calculations for $T/T_K =0.2$ (green solid line, bottom x-axis) and $T/T_K =0.4$ (red dashed line, top x-axis). Note that the experimental data in Fig.~2(b) can be read off either top or bottom axes, since $T_K$ is different for the experimental data when comparing with the two theoretical traces. For $T/T_K =0.20$, we find very good agreement over the entire bias range, and extract $T_K = 452\pm50$~mK. Increasing $T/T_K$ to 0.4 (red dashed line, top x-axis) fits the experiment well for $eV_{SD}/k_BT_K <0.4$ but deviates significantly at higher biases, indicating that the ratio $T/T_K=0.4$ is too high. The error in  $T_K$ is estimated by fitting the experimental trace to the two traces on each side of the optimal ratio and taking the average of the extracted $T_K$.

\begin{figure}
\includegraphics[width = 8.5cm]{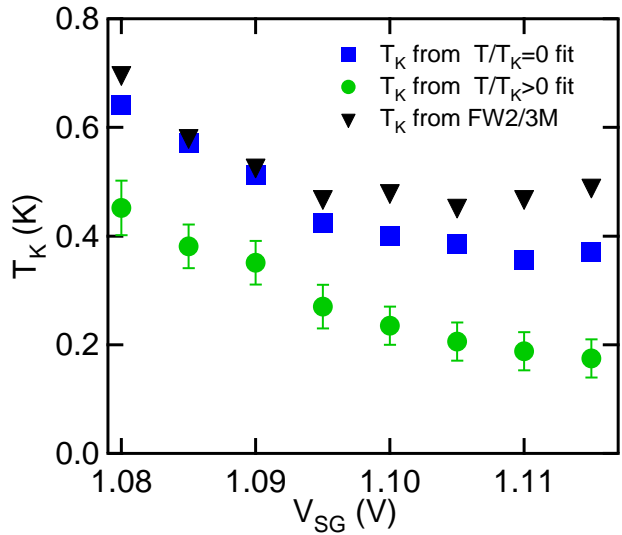}
\caption{\label{fig4}  (Color online) The  values of $T_K$ extracted from the fit of experimental data to  $T=0$ calculations  (blue squares),  $T/T_K >0$ calculations (green circles) and FW2/3M (black triangles) as a function of $V_{SG}$. }
\end{figure}

We perform the same fitting procedure for different side gate voltages, as shown in Fig.~3. We consistently get a better fit of experiment to theory when using calculations for $T/T_K > 0$ (solid green lines, bottom x-axis), compared to $T/T_K = 0$ (dashed blue lines, top x-axis). Using the values of $T/T_K$ and $T_K$ obtained  from Figs.~2(b) and 3 we can estimate the hole temperature $T$ for each of the experimental traces. We obtain $T=85 \pm 5$~mK consistent with the temperature obtained by analysing Shubnikov de Haas oscillations from a GaAs two dimensional hole system performed in the same setup.

In Fig.~4 we compare the values of $T_K$ obtained from fitting to the  $T/T_K = 0$ and $T/T_K > 0$ theory with those obtained from the peak width at $\frac{2}{3}$ maximum. From the shape of the universal $G'(V_{SD},T_K)/G_0$ conductance trace obtained from the $T=0$ theory (blue line in Fig.~1(b)), it can easily be seen that $G'/G_0=0.67$ occurs at $eV_{SD}=T_K$. Therefore the values of $T_K$ obtained from FW2/3M (black triangles) and the complete $T/T_K = 0$ fit (blue squares) in Fig.~4 should coincide as long as $T/T_K$ is small. We can see in Fig.~4  that both traces follow the same trend, although the two traces separate from each other at larger $V_{SG}$, where $T_K$ is small. This is because the $T = 0$ theory does not fit the experimental data well in this regime (see Fig.~3). Furthermore, both FW2/3M
and the $T/T_K=0$ fit significantly overestimate the value of the real $T_K$ obtained from fitting to the finite $T$ theory (green circles). This shows that the zero bias peak width can not be used as an accurate estimate of the Kondo temperature at finite measurement temperatures.

To conclude, we have measured the zero bias peak in differential conductance in a hole quantum dot. We have scaled the experimental data and compared it to  real time renormalization group calculations of differential conductance as a function of source-drain bias in the limit of zero temperature $T$ and at finite temperatures. Our experimental data show  deviations from the  $T=0$ calculations at  low bias and are in  very good agreement with the finite $T$ calculations.  The Kondo temperature $T_K$ obtained from fitting the data to finite $T$ calculations is significantly lower than that extracted from the peak width at $\frac{2}{3}$ maximum  and $T=0$ calculations.

This work was funded by the Australian Research Council through the DP, FT, APF and DORA schemes. Three co-authors (F.R, M.P., and
H.S.) acknowledge the financial support from the DFG via the project FOR~723.


\begin{thebibliography}{99}

\bibitem{Kondo64} J. Kondo, Prog. Theor. Phys. {\bf 32}, 37 (1964).

\bibitem{Haas33} W. J. de Haas, J. de Boer and G. J. Van den Berg, Physica {\bf 1}, 1115 (1933).

\bibitem{MadhavanScience98} V. Madhavan, W. Chen, T. Jamneala,  M. F. Crommie and  N. S. Wingreen,  Science {\bf 280}, 567 (1998).

\bibitem{GoldhaberNat98} D. Goldhaber-Gordon, H. Shtrikman, D. Mahalu, D. Abusch-Magder, U. Meirav and M. A. Kastner, Nature {\bf 391}, 156 (1998).

\bibitem{CronenwettScience98} S. M. Cronenwett, T. H. Oosterkamp and L. P. Kouwenhoven, Science {\bf 281}, 540 (1998).

\bibitem{KouwenhovenPhysWorld} L. P. Kouwenhoven and L. I. Glazman, Phys. World  {\bf 14}, 33 (2001).

\bibitem{GoldhaberPRL98} D. Goldhaber-Gordon, J. G\"{o}res, M. A. Kastner, Hadas Shtrikman, D. Mahalu, and U. Meirav, Phys. Rev. Lett. {\bf 81}, 5225 (1998).

\bibitem{PletyukhovPRL12} M. Pletyukhov and H. Schoeller, Phys. Rev. Lett. {\bf 108}, 260601 (2012).

\bibitem{KretininPRB12} A. V. Kretinin, H. Shtrikman, and D. Mahalu, Phys. Rev. B {\bf 85}, 201301(R) (2012).

\bibitem{Reininghaus12} F. Reininghaus, M. Pletyukhov and H. Schoeller, {\it in preparation} (2013).

\bibitem{ChenNJP10}  J. C. H. Chen, O. Klochan, A. P. Micolich, A. R. Hamilton, T. P. Martin, L. H. Ho, U. Z\"{u}licke, D. Reuter and A. D. Wieck,  New J. Phys. {\bf 12}, 033043 (2010).

\bibitem{KlochanPRL11} O. Klochan, A. P. Micolich, A. R. Hamilton, K. Trunov, D. Reuter, and A. D. Wieck, Phys. Rev. Lett. {\bf 107}, 076805 (2011).


\bibitem{EnsslinNP06} K. Ensslin, Nat. Phys. {\bf 2}, 587 (2006).

\bibitem{SfigakisPRL08} F. Sfigakis,  C. J. B.  Ford,  M. Pepper,   M.  Kataoka,  D. A. Ritchie,  M. Y. Simmons, Phys. Rev. Lett. {\bf 100}, 026807 (2008).

\bibitem{KlochanAPL06} O. Klochan, W. R. Clarke, R. Danneau, A. P. Micolich, L. H. Ho, A. R. Hamilton, K. Muraki and Y. Hirayama, Appl. Phys. Lett. {\bf 89}, 092105 (2006).

\bibitem{bias} A series resistance of $\approx 30.5$~k$\Omega$ due to ohmic contacts is used to calculate the voltage drop across the dot.

\bibitem{CronenwettPRL02} S. M. Cronenwett,  H. J. Lynch, D.  Goldhaber-Gordon,   L. P. Kouwenhoven, C. M. Marcus, K.  Hirose,  N. S. Wingreen  and V. Umansky, Phys. Rev. Lett. {\bf 88}, 226805 (2002).

\bibitem{SarkozyPRB09} S. Sarkozy,    F. Sfigakis,    K. Das Gupta,  I. Farrer,   D. A. Ritchie,   G. A. C. Jones  and M. Pepper, Phys. Rev. B {\bf 79}, 161307(R) (2009).

\bibitem{odd} If the dot occupancy were even, the Kondo effect could also arise due to single-triplet degeneracy which can be tuned either by magnetic field~\cite{SasakiNature00} or confinement potential~\cite{KoganPRL03}. However, the even Kondo effect is characterized by nonmonotonic dependence of the ZBP in magnetic field and temperature~\cite{GrangerPRB05}. We do not observe these signatures. Moreover in our device, the splitting of the ZBP in magnetic field is double the Zeeman energy, which is only expected for odd dot occupancy~\cite{KlochanPRL11}.

\bibitem{SasakiNature00} S. Sasaki, S. De Franceschi, J. M. Elzerman, W. G. van der Wiel, M. Eto, S. Tarucha and L. P. Kouwenhoven, Nature {\bf 405}, 764 (2000).

\bibitem{KoganPRL03} A. Kogan, S. Amasha, D. Goldhaber-Gordon, G. Granger, M. A. Kastner, and H. Shtrikman, Phys. Rev. Lett. {\bf 67}, 113309 (2003).

\bibitem{GrangerPRB05} G. Granger, M. A. Kastner, I. Radu, M. P. Hanson, and A. C. Gossard, Phys. Rev. B {\bf 72}, 165309 (2005).

\bibitem{KretininPRB11} A. V. Kretinin, H. Shtrikman, D. Goldhaber-Gordon, M. Hanl, A. Weichselbaum, J. von Delft, T. Costi, and D. Mahalu, Phys. Rev. B {\bf 84}, 245316 (2011).

\bibitem{MartinMorenoJPCM92} L. Martin-Moreno, J. T. Nicholls, N. K. Patel and M. Pepper, J. Phys.: Condens. Matter {\bf 4}, 1323 (1992).

\bibitem{PustilnikJPC04} M. Pustilnik and L. Glazman, J. Phys.: Condens. Matter {\bf 16}, R513 (2004).


\end{thebibliography}
\end{document}